\documentclass[12pt,a4paper]{article}  
\usepackage{a4wide}
\usepackage{graphicx}
\usepackage{subfigure}
\usepackage[centertags]{amsmath}
\usepackage{amssymb}
\usepackage{cite}
\usepackage{epsfig}
\usepackage{lscape}
\usepackage{longtable}
\usepackage{scrtime}
\usepackage{epic}
\usepackage{eepic}
\usepackage{color}

\usepackage{dsfont}
\usepackage{enumerate}
\usepackage{theorem}
\theoremstyle{plain}
\theorembodyfont{}

\definecolor{brown}{rgb}{0.75,0.37,0.00}

\newcommand{\bs}[1]{\boldsymbol{#1}}
\newcommand{\bsb}[1]{\boldsymbol{\overline{#1}}}
\newcommand{\fr}[2]{\frac{#1}{#2}}

\newcommand{\2}{\frac{1}{2}}
\newcommand{\4}{\frac{1}{4}}
\newcommand{\8}{\frac{1}{8}}

\begin{document}  

\vskip 2cm

\begin{center}  
{\huge Exploring the SO(32) Heterotic String}
\vspace*{5mm} \vspace*{1cm}   
\end{center}  
\vspace*{5mm} \noindent  
\vskip 0.5cm  
\centerline{\bf Hans Peter Nilles$^a$, Sa\'ul Ramos-S\'anchez$^a$, Patrick 
  Vaudrevange$^a$, Ak\i{}n Wingerter$^b$}
\vskip 1cm
\centerline{$^a$\em Physikalisches Institut, Universit\"at Bonn}
\centerline{\em Nussallee 12, D-53115 Bonn, Germany}
\vskip 0.5cm
\centerline{$^b$ \em Department of Physics, The Ohio State University}
\centerline{\em 191 W.\ Woodruff Ave., Columbus, OH 43210, USA}
\vskip2cm
  
\centerline{\bf Abstract}  
\vskip .3cm

We give a complete classification of $\mathbb{Z}_N$ orbifold compactification
of the heterotic SO(32) string theory and show its potential for
realistic model building. The appearance of spinor representations
of SO(2$n$) groups is analyzed in detail. We conclude that the 
heterotic SO(32) string constitutes an interesting 
part of the string landscape both in view of model constructions
and the question of heterotic-type I duality.

\vskip .3cm

\newpage

\section{Introduction}

String theory might provide a framework to describe all particle physics
phenomena. Still we do not know how to derive the standard model of strong
and electro-weak interactions from first principles. Apparently many roads
seem to be possible: the so-called landscape of string vacua. 
Progress might
be made by exploring this landscape in detail 
to understand possible phenomenological
patterns that might be mapped to experimental observations.
 Such patterns might
include concepts like supersymmetry, grand unification
and extra dimensions.

In the present paper we would like to explore the heterotic SO(32) string
theory and its suitability for model building. There has been less effort
spent on the SO(32) theory than its $\text{E}_8\times\text{E}_8'$ brother, which was 
considered as the prime
candidate initially. A detailed analysis of the SO(32) theory shows,
however, that model building within this framework could be as exciting as
in the $\text{E}_8\times\text{E}_8'$ case.

An additional motivation to consider the SO(32) 
heterotic theory is the
exploration of the conjectured duality to the SO(32) 
type I string theory~\cite{Polchinski:1995df}. This
might prove useful to understand connections between heterotic 
model constructions
and those based on type II orientifolds.

Our analysis considers the orbifold compactification\cite{Dixon:1985jw} of the SO(32) heterotic
theory\footnote{For earlier work on SO(32) heterotic string orbifolds, see ref.~\cite{Giedt:2003an,Choi:2004wn}.}, as it combines the complexity of Calabi-Yau compactification with
the calculability of torus compactification. 
Many phenomenological
properties find a geometric explanation in this framework~\cite{Forste:2004ie, Forste:2005rs, Forste:2005gc}. 
We derive a complete classification
of the four-dimensional heterotic SO(32) orbifold constructions. This is
necessary as previous attempts to do so have been 
found to be incomplete. We explain the
subtleties  of the construction and give a 
detailed presentation of the $\mathbb{Z}_4$-orbifold.
The remaining cases are given in detail on a web page~\cite{SO32:webseite} that will be made
available to the public.

Having achieved this goal of classification we explore  properties that
might be important for explicit model building. One aspect e.g. 
is the question of the appearance of 
spinor representations of SO(2$n$) gauge groups (with $n=5,6,7$). 
Spinors of SO(10) \cite{Georgi:1974my,Fritzsch:1974nn} e.g. would be very suitable for a 
description of families of quarks
and leptons, as argued in ref.~\cite{Nilles:2004ej}. 
In addition, the appearance of these
spinors might be relevant to understand the nature of the 
heterotic-type~I duality in four space time dimensions~\cite{Angelantonj:1996uy,Kakushadze:1997wx,Lalak:1999bk,Blumenhagen:2005pm}.

Using this information we provide a few explicit examples of 3-family models
in this framework to illustrate the ease with which such models can be 
constructed. One example is obtained even in the absence of Wilson lines.
Our results can be used as a starting point for a full classification of
models including Wilson lines, the inclusion of which is, 
however, beyond the scope of this paper.
Nonetheless, some useful patterns of possible spectra can 
be deduced from our
results with the concept of fixed-point equivalent models~\cite{Gmeiner:2002es}. 
It thus appears that
the heterotic SO(32) theory is a fertile part of the string 
theory landscape.

The paper is organized as follows. In section \ref{subsec:general_classification} we present the
strategy to classify all orbifolds of the SO(32) heterotic
string. In section \ref{subsec:Z4_classification} and \ref{subsec:ineq_models} we illustrate the method for the
$\mathbb{Z}_4$ orbifold explicitly and give the list of models for the
$\mathbb{Z}_N$ orbifolds. Section 3 is devoted to the discussion of
the spinorial representations of SO(2$n$) gauge groups for
various $n$. Two explicit examples of 3-family models will be
presented in section 4, followed by concluding remarks in section 5.
Some technical details and tables are given in the appendices.

\section{Classification of Orbifolds}
\label{sec:classification_of_orbifolds}

\subsection{Classification of $\boldsymbol{\text{SO}(32)}$ Orbifold Models}
\label{subsec:general_classification}

To introduce the relevant notation \cite{Ibanez:1987pj,Forste:2004ie} and to set the stage for the
following calculations, we briefly summarize some of the concepts in orbifold constructions, before
proceeding to describe the classification of inequivalent models.\\[-1ex]

An orbifold is defined to be the quotient of a torus\footnote{In a more general context, an orbifold is defined to be the quotient of a manifold by a discrete symmetry.} by a discrete set of its isometries, called the {\it point group} $P$. Modular invariance requires the action of the point group to be accompanied by a corresponding action $G$ ({\it gauge twisting group}) on the 16 gauge degrees of freedom:
\begin{equation}
\mathcal{O} = T^6 \big{/}P \otimes T^{16} \big{/} G
\label{eq:define_orbifold_by_point_group}
\end{equation}
Modular invariance and the homomorphism property of the gauge embedding $P \hookrightarrow G$ put further
restrictions on $G$, which will be discussed later. Consistency with ten-dimensional anomaly cancellation
requires $T^{16}$ to be an even, integral and self-dual lattice. In 16 dimensions, there are only 2
admissible choices, namely the root lattice of $\text{E}_8\times\text{E}_8'$ and the weight lattice of
$\text{Spin}(32)/\mathbb{Z}_2$. Here, we focus our attention on the latter case.\\[-1ex]

The representations of $\text{Spin}(32)$ fall into 4 conjugacy classes, corresponding to the adjoint, vector, spinor and conjugate spinor representation, respectively \cite{Green:1987sp,Slansky:1981yr}. Two representations are said to be conjugate, if their weight vectors differ by an element of the root lattice $\Lambda_R$. Consequently, the weight lattice $\Lambda_W$ can be written as the sum of 4 disjoint sublattices, given by the highest weight of the respective representation modulo $\Lambda_R$. By $\text{Spin}(32)/\mathbb{Z}_2$ we shall understand the symmetry corresponding to the adjoint and spinor conjugacy classes, and denote the respective lattice by $\Lambda_{\text{Spin}(32)/\mathbb{Z}_2}$.\\[-1ex]

The action of $G$ on $T^{16}$ can be described as a shift $X_L\mapsto X_L+V$ \cite{Dixon:1986jc}, which induces the transformations
\begin{equation}
\sigma_V(H_i) = H_i, \qquad \sigma_V(E_{\alpha}) = \exp\left(2\pi i \,\alpha\cdot V \right) E_{\alpha}
\label{eq:action_shift_on_operators}
\end{equation}
on the Cartan generators and step operators of $\text{SO}(32)$, and these transformations clearly
describe an automorphism of the algebra\footnote{Note that the group $\text{SO}(32)$ and its covering
  group $\text{Spin}(32)$ share the same algebra.}. The automorphisms of semi-simple Lie algebras have been
classified \cite{Kac:1969xxx1}, and it is straightforward to obtain the corresponding shifts, as we will now describe.\\[-1ex]

\begin{figure}
\begin{center}
\setlength{\unitlength}{0.75mm} 
\begin{picture}(130,22) 

\put(10,19){\circle{2}}
\put(10,3){\circle{2}}
\put(18,11){\circle{2}}
\put(26,11){\circle{2}}
\put(34,11){\circle{2}}
\put(42,11){\circle{2}}
\put(50,11){\circle{2}}
\put(58,11){\circle{2}}
\put(66,11){\circle{2}}
\put(74,11){\circle{2}}
\put(82,11){\circle{2}}
\put(90,11){\circle{2}}
\put(98,11){\circle{2}}
\put(106,11){\circle{2}}
\put(114,11){\circle{2}}
\put(122,19){\circle{2}}
\put(122,3){\circle{2}}

\put(10.7071,18.2929){\line(1,-1){6.55}} 
\put(10.7071,3.7071){\dashline{2}(0,0)(6,6)}
\put(19,11){\line(1,0){6}} 
\put(27,11){\line(1,0){6}} 
\put(35,11){\line(1,0){6}} 
\put(43,11){\line(1,0){6}} 
\put(51,11){\line(1,0){6}} 
\put(59,11){\line(1,0){6}} 
\put(67,11){\line(1,0){6}} 
\put(75,11){\line(1,0){6}} 
\put(83,11){\line(1,0){6}} 
\put(91,11){\line(1,0){6}} 
\put(99,11){\line(1,0){6}} 
\put(107,11){\line(1,0){6}} 
\put(114.7071,11.7071){\line(1,1){6.55}} 
\put(114.7071,10.2929){\line(1,-1){6.55}} 

\put(8,-1){${\scriptstyle\alpha_0}$}
\put(8,15){${\scriptstyle\alpha_1}$}
\put(16,6){${\scriptstyle\alpha_2}$}
\put(24,6){${\scriptstyle\alpha_3}$}
\put(32,6){${\scriptstyle\alpha_4}$}
\put(40,6){${\scriptstyle\alpha_5}$}
\put(48,6){${\scriptstyle\alpha_6}$}
\put(56,6){${\scriptstyle\alpha_7}$}
\put(64,6){${\scriptstyle\alpha_8}$}
\put(72,6){${\scriptstyle\alpha_9}$}
\put(79,6){${\scriptstyle\alpha_{10}}$}
\put(87,6){${\scriptstyle\alpha_{11}}$}
\put(95,6){${\scriptstyle\alpha_{12}}$}
\put(103,6){${\scriptstyle\alpha_{13}}$}
\put(111,6){${\scriptstyle\alpha_{14}}$}
\put(121,-1){${\scriptstyle\alpha_{15}}$}
\put(121,15){${\scriptstyle\alpha_{16}}$}

\put(9,5){${\scriptscriptstyle 1}$}
\put(9,21){${\scriptscriptstyle 1}$}
\put(17,13){${\scriptscriptstyle 2}$}
\put(25,13){${\scriptscriptstyle 2}$}
\put(33,13){${\scriptscriptstyle 2}$}
\put(41,13){${\scriptscriptstyle 2}$}
\put(49,13){${\scriptscriptstyle 2}$}
\put(57,13){${\scriptscriptstyle 2}$}
\put(65,13){${\scriptscriptstyle 2}$}
\put(73,13){${\scriptscriptstyle 2}$}
\put(81,13){${\scriptscriptstyle 2}$}
\put(89,13){${\scriptscriptstyle 2}$}
\put(97,13){${\scriptscriptstyle 2}$}
\put(105,13){${\scriptscriptstyle 2}$}
\put(113,13){${\scriptscriptstyle 2}$}
\put(121,5){${\scriptscriptstyle 1}$}
\put(121,21){${\scriptscriptstyle 1}$}

\end{picture} 
\end{center}
\caption{Extended Dynkin diagram of $\text{SO}(32)$ and the associated Ka\v{c} labels.}
\label{fig:extended_dynkin_diagram_of_SO32}
\end{figure}
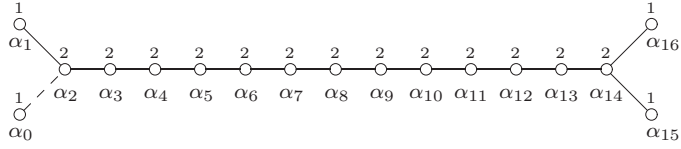

\subsubsection*{Automorphisms of $\boldsymbol{\text{SO}(32)}$}

To this end, consider the extended Dynkin diagram of SO(32) given in fig.~\ref{fig:extended_dynkin_diagram_of_SO32}. The numbers which have been adjoined to the nodes are the Ka\v{c} labels $k_i$, which are by definition the expansion coefficients of the highest root $\alpha_H$ in terms of the simple roots, i.e.~
\begin{equation}
\alpha_H = k_1 \alpha_1 + \ldots + k_\ell \alpha_\ell,
\label{eq:def_kac_labels}
\end{equation} 
where $\ell$ is the rank\footnote{Clearly, $\ell=16$ in our case, but for the time being, we want to keep the discussion general.} of the algebra. For convenience, the Ka\v{c} label of the {\it most negative root} $\alpha_0 \equiv -\alpha_H$ is set to $k_0 = 1$. Then, by a theorem due to Ka\v{c} \cite{Kac:1969xxx1}, all order-N automorphisms of an algebra up to conjugation are given by
\begin{equation}
\sigma_{s,m}(E_{\alpha_j}) = \mu\,\exp\left(2\pi i s_j/N\right) E_{\alpha_j}, \quad j = 0, \ldots, \ell,
\label{eq:action_kac_on_operators}
\end{equation}
where the sequence $s = (s_0, \ldots, s_\ell)$ may be chosen arbitrarily subject to the conditions that the $s_i$ are non-negative, relatively prime integers and
\begin{equation}
N = m\sum_{i=0}^{\ell} k_i s_i.
\label{eq:order_N_intermsof_s_i}
\end{equation}
Hereby, $\mu$ is an automorphism of the Dynkin diagram, and $m$ is the smallest integer such that
$(\sigma_{s,m})^m$ is inner. Since in this context we are only interested in inner automorphisms, we set
$\mu = \mathds{1}$ and $m=1$. Furthermore, it should be noted that two automorphisms $\sigma_{s}$ and
$\sigma_{s'}$ are conjugate if and only if the sequence $s$ can be transformed into the sequence $s'$ by
a symmetry of the extended diagram. In section \ref{subsec:Z4_classification}, we will encounter an
interesting example which shows that two such automorphisms of $\text{SO}(32)$ must not be identified.
\bigskip 

\subsubsection*{The Shift Vector}

To derive the shift vector corresponding to a given automorphism is now particularly easy. Comparing eq.~(\ref{eq:action_shift_on_operators}) to eq.~(\ref{eq:action_kac_on_operators}), it immediately follows that
\begin{equation}
\alpha_i \cdot V = \frac{s_i}{N}, \quad i=1, \ldots, \ell,
\label{eq:lineq_for_V}
\end{equation}
for the $\ell$ linearly independent roots $\alpha_i$. Expanding $V$ in terms of the dual simple roots and substituting this expression in the previous equation gives
\begin{equation}
V = \frac{1}{N} \left( s_1 \alpha_1^* + \ldots + s_\ell \alpha_\ell^* \right),
\label{eq:V_in_Dynkin_basis}
\end{equation}
i.e. the integers $s_i$ divided by the order $N$ are the Dynkin labels of $V$. It is checked by a direct calculation that this $V$ also gives the correct transformation for the step operator corresponding to the most negative root.\\[-1ex]

Determining the unbroken gauge group is now particularly simple. Looking at
eq.~(\ref{eq:V_in_Dynkin_basis}) we see that in the extended Dynkin diagram, the root $\alpha_i$
($i=0,\ldots,\ell$) is projected out, if and only if the coefficient $s_i$ in
eq.~(\ref{eq:order_N_intermsof_s_i}) does not
vanish. To calculate the spectrum of the orbifold, we need an explicit expression for the shift vector
$V$, which is easily obtained once the simple roots and their duals are given. For a standard choice of
roots, see e.g.~ref.~\cite{Green:1987sp}.

\subsubsection*{Restrictions on the Shift Vector}

Not every shift vector $V$ which describes an automorphism of the algebra is an admissible choice for model construction. For a twist $\theta \in P$ of order N,\, $\theta^N = \mathds{1}$ implies that $N\,V$ should act as the identity on $T^{16}$, and hence, from the self-duality of the lattice, it immediately follows that
\begin{equation}
N\,V \in \Lambda_{\text{Spin}(32)/\mathbb{Z}_2}.
\label{eq:gauge_embedding_hom_property}
\end{equation}
By eq.~(\ref{eq:V_in_Dynkin_basis}), $N\,V$ is only guaranteed to lie in the weight lattice $\Lambda_W$, so that some of the shift vectors will be ruled out. \\[-1ex]

For the partition function of a $\mathbb{Z}_N$ orbifold to be modular invariant, the relation
\begin{equation}
N\left( V^2 - v^2 \right) = 0 \text{ mod } 2
\label{eq:mod_inv_general}
\end{equation}
has to be satisfied \cite{Dixon:1986jc}, where $v$ is a 3-dimensional vector describing the action of the twist on the complexified, compact coordinates. This condition severely restricts the number of shift vectors which can be used in constructing orbifold models.\\[-1ex]

From eq.~(\ref{eq:gauge_embedding_hom_property}) it is clear that for a given order $N$ of the twist $\theta$, all shifts $V$ of order $M$ are also admissible, as long as $M$ divides $N$. In principle, we could determine the admissible shifts for each $M$ separately, but a more practical approach is to run through the outlined procedure for $N$, dropping the condition on the relative-primeness of the sequence $s = (s_0, \ldots, s_\ell)$, see the remarks preceding eq.~(\ref{eq:order_N_intermsof_s_i}). In the cases where $s$ is not relatively prime and the common divisor can be cancelled out from both the numerator and the denominator in eq.~(\ref{eq:V_in_Dynkin_basis}), the order of the shift is some $M$ which is smaller than $N$.\\[-1ex]

We will illustrate the outlined methods using the $\mathbb{Z}_4$ orbifold in section \ref{subsec:Z4_classification}.

\subsection{The $\boldsymbol{\mathbb{Z}_4}$ Orbifold}
\label{subsec:Z4_classification}

\subsubsection*{Classification}
We shall use the method presented in section~\ref{subsec:general_classification} to compute all
admissible shifts for the $\mathbb{Z}_4$ orbifold. For $N=4$, there are 256 different vectors
$s=(s_0,\ldots,s_{16})$, which satisfy eq.~(\ref{eq:order_N_intermsof_s_i}) with m=1 and the Ka\v{c} labels
$k_i$ given in fig.~\ref{fig:extended_dynkin_diagram_of_SO32}. We express the corresponding shift
vectors using eq.~(\ref{eq:V_in_Dynkin_basis}) and a standard choice of roots\cite{Green:1987sp}. Of 
these shift vectors, 134 satisfy the first
restriction eq.~(\ref{eq:gauge_embedding_hom_property}) for admissible orbifold shifts and only 30 are
left when we impose the modular invariance requirement, given by eq.~(\ref{eq:mod_inv_general}) with the
twist $v=\tfrac{1}{4}(1,1,\text{-}2)$. Considering two shifts to be inequivalent if their spectra are
different, we find only 16 inequivalent shift vectors in the $\mathbb{Z}_4$ orbifold. These are all
possible shifts one can obtain.

\subsubsection*{Anomalies}
The 16 inequivalent shift vectors, their corresponding gauge groups and spectra are listed in
table~\ref{tab:Z4_classification} of appendix~\ref{sec:Z4_models}. We have denoted the anomalous
$\text{U}_1$ factors by $\text{U}_{1A}$. As a cross-check for our calculations, we have verified the
following conditions for anomaly cancellation
\begin{equation}
\label{eq:Anomaly_conditions}
\frac{1}{24}\text{Tr}\,Q_i = \frac{1}{6 |t_i|^2}\text{Tr}\,Q_i^3 = \frac{1}{2}\text{Tr}\,lQ_i = 
   \left\{ \begin{array}{ll}\frac{1}{2|t_j|^2}\text{Tr}\,Q_j^2 Q_A \neq 0 \quad & \text{if } i = A,\,j \neq A \\ 0 &
         \text{otherwise}\end{array} \right.
\end{equation}
where $l$ denotes the index of a given representation. Furthermore, $t_i$ is the generator of the i-th $\text{U}_1$
factor that defines the charge $Q_i$ as:
\begin{equation}
Q_i |p_{sh}\rangle_L = (t_i \cdot p_{sh}) |p_{sh}\rangle_L\text{ ,}
\end{equation}
where $p_{sh}$ is the shifted $\text{Spin}(32)/\mathbb{Z}_2$ lattice vector. In the case when eq.~(\ref{eq:Anomaly_conditions}) does not vanish, these conditions guarantee that the anomalous
$\text{U}_1$ is cancelled by the generalized Green-Schwarz mechanism~\cite{Green:1984sg,Witten:1984dg,Dine:1987xk,Sagnotti:1992qw,Berkooz:1996iz},~\cite{Blumenhagen:2005pm}.

\subsubsection*{Discussion of the Results}
A detailed list including the spectra of all
$\mathbb{Z}_4$ orbifold models is given in table~\ref{tab:Z4_classification} of
appendix~\ref{sec:Z4_models}. In particular, in the second column of this table we compare our results to those
previously obtained
in ref.~\cite{Choi:2004wn}. In ref.~\cite{Choi:2004wn} the shift vectors are separated into two classes. The
so-called {\em vectorial} shifts in the $\mathbb{Z}_4$ orbifold are those whose entries have a maximal
denominator of 4, whereas all entries of the {\em spinorial} shifts have a denominator of 8 and an odd
numerator. Using these definitions, 12 of our shifts are vectorial and 4 are
spinorial. 

Our result differs in some ways from that of the ref.~\cite{Choi:2004wn}. 
Some multiplicities
of states and $\text{U}(1)$ charges are different from our findings and cannot be related by a change of
basis in the $\text{U}(1)$ directions. For the models in question, ref.~\cite{Choi:2004wn} does not
fulfill the anomaly cancellation conditions, eq.~(\ref{eq:Anomaly_conditions}).

Additionally, we find 16 inequivalent models whereas one can obtain only 10 inequivalent shifts with the
general method proposed in ref.~\cite{Choi:2004wn}. This discrepancy is related to two problems.
One is that by using the ansatz for a spinorial shift proposed in ref.~\cite{Choi:2004wn} and the weight lattice 
as given in section \ref{subsec:general_classification}, one cannot
obtain any of the four spinorial shifts we found in a direct manner for $\mathbb{Z}_4$. 
In appendix~\ref{sec:ansatz} we give an alternative ansatz for the form of any shift of $\mathbb{Z}_N$ orbifolds in
the SO(32) heterotic string. Yet a classification based on this ansatz
is more time consuming than the method presented in section~\ref{subsec:general_classification}.

The second problem is that the shift vectors V$^{(4)}$ and V$^{(12)}$ of our vectorial shifts are not
listed in ref.~\cite{Choi:2004wn}. Here, V$^{(i)}$ denotes the shift vector corresponding to the model
number $i$ of table~\ref{tab:Z4_classification}. The reason can be traced back to comparing, for instance, the shifts V$^{(3)}$
and V$^{(4)}$ of our list. Since both shifts generate the same unbroken gauge
group in four dimensions and the same matter representations in the untwisted and second twisted sectors,
one might be tempted to consider them to be equivalent. But the matter content of the first twisted
sector is different, therefore, the two shifts lead to different models.

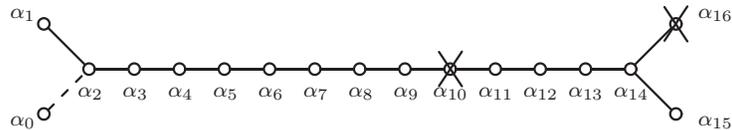
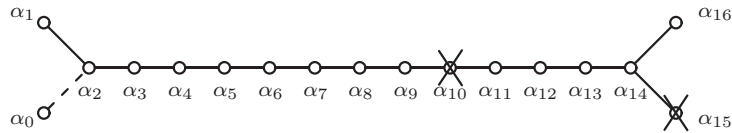
\begin{figure}[!h]
\begin{center}

\subfigure[Breaking due to V$^{(3)}$]{
\setlength{\unitlength}{0.75mm} 
\begin{picture}(130,22) 
\thicklines
\put(10,19){\circle{2}}
\put(10,3){\circle{2}}
\put(18,11){\circle{2}}
\put(26,11){\circle{2}}
\put(34,11){\circle{2}}
\put(42,11){\circle{2}}
\put(50,11){\circle{2}}
\put(58,11){\circle{2}}
\put(66,11){\circle{2}}
\put(74,11){\circle{2}}
\put(82,11){\circle{2}}
\put(90,11){\circle{2}}
\put(98,11){\circle{2}}
\put(106,11){\circle{2}}
\put(114,11){\circle{2}}
\put(122,19){\circle{2}}
\put(122,3){\circle{2}}

\put(10.7071,18.2929){\line(1,-1){6.55}} 
\put(10.7071,3.7071){\dashline{2}(0,0)(6,6)}
\put(19,11){\line(1,0){6}} 
\put(27,11){\line(1,0){6}} 
\put(35,11){\line(1,0){6}} 
\put(43,11){\line(1,0){6}} 
\put(51,11){\line(1,0){6}} 
\put(59,11){\line(1,0){6}} 
\put(67,11){\line(1,0){6}} 
\put(75,11){\line(1,0){6}} 
\put(83,11){\line(1,0){6}} 
\put(91,11){\line(1,0){6}} 
\put(99,11){\line(1,0){6}} 
\put(107,11){\line(1,0){6}} 
\put(114.7071,11.7071){\line(1,1){6.55}} 
\put(114.7071,10.2929){\line(1,-1){6.55}} 

\put(80,14){\line(2,-3){4.1}}
\put(80,8){\line(2,3){4.1}}
\put(120,22){\line(2,-3){4.1}}
\put(120,16){\line(2,3){4.1}}

\put(4,1){${\scriptstyle\alpha_0}$}
\put(4,20){${\scriptstyle\alpha_1}$}
\put(16,6){${\scriptstyle\alpha_2}$}
\put(24,6){${\scriptstyle\alpha_3}$}
\put(32,6){${\scriptstyle\alpha_4}$}
\put(40,6){${\scriptstyle\alpha_5}$}
\put(48,6){${\scriptstyle\alpha_6}$}
\put(56,6){${\scriptstyle\alpha_7}$}
\put(64,6){${\scriptstyle\alpha_8}$}
\put(72,6){${\scriptstyle\alpha_9}$}
\put(79,6){${\scriptstyle\alpha_{10}}$}
\put(87,6){${\scriptstyle\alpha_{11}}$}
\put(95,6){${\scriptstyle\alpha_{12}}$}
\put(103,6){${\scriptstyle\alpha_{13}}$}
\put(111,6){${\scriptstyle\alpha_{14}}$}
\put(126,1){${\scriptstyle\alpha_{15}}$}
\put(126,20){${\scriptstyle\alpha_{16}}$}

\end{picture} 
}

\subfigure[Breaking due to V$^{(4)}$]{

\setlength{\unitlength}{0.75mm} 
\begin{picture}(130,22) 
\thicklines
\put(10,19){\circle{2}}
\put(10,3){\circle{2}}
\put(18,11){\circle{2}}
\put(26,11){\circle{2}}
\put(34,11){\circle{2}}
\put(42,11){\circle{2}}
\put(50,11){\circle{2}}
\put(58,11){\circle{2}}
\put(66,11){\circle{2}}
\put(74,11){\circle{2}}
\put(82,11){\circle{2}}
\put(90,11){\circle{2}}
\put(98,11){\circle{2}}
\put(106,11){\circle{2}}
\put(114,11){\circle{2}}
\put(122,19){\circle{2}}
\put(122,3){\circle{2}}

\put(10.7071,18.2929){\line(1,-1){6.55}} 
\put(10.7071,3.7071){\dashline{2}(0,0)(6,6)}
\put(19,11){\line(1,0){6}} 
\put(27,11){\line(1,0){6}} 
\put(35,11){\line(1,0){6}} 
\put(43,11){\line(1,0){6}} 
\put(51,11){\line(1,0){6}} 
\put(59,11){\line(1,0){6}} 
\put(67,11){\line(1,0){6}} 
\put(75,11){\line(1,0){6}} 
\put(83,11){\line(1,0){6}} 
\put(91,11){\line(1,0){6}} 
\put(99,11){\line(1,0){6}} 
\put(107,11){\line(1,0){6}} 
\put(114.7071,11.7071){\line(1,1){6.55}} 
\put(114.7071,10.2929){\line(1,-1){6.55}} 

\put(80,14){\line(2,-3){4.1}}
\put(80,8){\line(2,3){4.1}}
\put(120,6){\line(2,-3){4.1}}
\put(120,0){\line(2,3){4.1}}

\put(4,1){${\scriptstyle\alpha_0}$}
\put(4,20){${\scriptstyle\alpha_1}$}
\put(16,6){${\scriptstyle\alpha_2}$}
\put(24,6){${\scriptstyle\alpha_3}$}
\put(32,6){${\scriptstyle\alpha_4}$}
\put(40,6){${\scriptstyle\alpha_5}$}
\put(48,6){${\scriptstyle\alpha_6}$}
\put(56,6){${\scriptstyle\alpha_7}$}
\put(64,6){${\scriptstyle\alpha_8}$}
\put(72,6){${\scriptstyle\alpha_9}$}
\put(79,6){${\scriptstyle\alpha_{10}}$}
\put(87,6){${\scriptstyle\alpha_{11}}$}
\put(95,6){${\scriptstyle\alpha_{12}}$}
\put(103,6){${\scriptstyle\alpha_{13}}$}
\put(111,6){${\scriptstyle\alpha_{14}}$}
\put(126,1){${\scriptstyle\alpha_{15}}$}
\put(126,20){${\scriptstyle\alpha_{16}}$}

\end{picture} 
}
\caption{Extended Dynkin diagram of SO(32) corresponding to the breaking due to the shifts V$^{(3)}$ and
  V$^{(4)}$ of the  $\mathbb{Z}_4$ orbifold.}
\label{fig:V3_V4_breaking}
\end{center}
\end{figure}

We have deeper reasons to argue that these two shifts are inequivalent. First, as illustrated in
fig.~\ref{fig:V3_V4_breaking}, the shift vectors come from two different breakings of the original SO(32)
gauge group in ten dimensions. Two different breakings may be
equivalent if one can transform the corresponding shifts into each other by adding lattice vectors
and applying automorphisms of the lattice. We can see that there is no lattice vector relating
the shifts V$^{(3)}$ and V$^{(4)}$. There exists an automorphism of $\text{SO}(32)$, which maps
V$^{(3)}$ onto V$^{(4)}$ up to a lattice vector, compare fig.~\ref{fig:V3_V4_breaking}.
It is not an automorphism of
$\text{Spin(32)}/\mathbb{Z}_2$, as it transforms the {\em spinor} conjugacy class of the lattice of
$\text{Spin(32)}/\mathbb{Z}_2$ into the {\em conjugate-spinor} class of Spin(32), which is not part of
$\text{Spin(32)}/\mathbb{Z}_2$.

In summary, this shows that the ansatz of ref.~\cite{Choi:2004wn} is incomplete. It leads only to 10 of
the 16 shift vectors in the $\mathbb{Z}_4$ orbifold. As we shall show in section~\ref{subsec:Z4_model}, one of
the missing shifts leads to a three-family model.

\subsection{The $\boldsymbol{\mathbb{Z}_N}$ Orbifold}
\label{subsec:ineq_models}

Using the method described in section~\ref{subsec:general_classification}, we have computed all
inequivalent models, which we do not list due to space limitations. All $\mathbb{Z}_N$ shifts, their
corresponding gauge groups and spectra are listed on our web page \cite{SO32:webseite}. 

\subsubsection*{Lists of Models}

In summary, there are 5141 $\mathbb{Z}_N$ orbifold models without Wilson lines. We have used the geometry 
of $\mathbb{Z}_N$ orbifolds as given in ref.~\cite{Kobayashi:1991rp}, and for the $\mathbb{Z}_8\text{-I}$ as 
given in ref.~\cite{Casas:1991ac}. On our web site \cite{SO32:webseite}, we provide for each model the
following details: 
\begin{itemize}
\item the twist $v$ and the 6 dimensional root lattice, which specifies the geometry,\\[-4ex]
\item the gauge shift $V$ and the corresponding gauge group,\\[-4ex]
\item the matter content, listed by sectors, including all $\text{U}_1$ charges, where we have denoted the anomalous one by $\text{U}_{1A}$.\footnote{More details are available from the authors upon request.}
\end{itemize}
For convenience, we have
implemented a search engine, with which one can choose models with a given gauge group.
As a side remark, this work can be seen as a contribution to the String Vacuum Project \cite{SVP} in the context of the heterotic string \cite{Dienes:2006ut}.

\subsubsection*{Discussion of the Results}
In table~\ref{tab:compareSO32andE8xE8} and table~\ref{tab:ineq_models}, we summarize our results. Our
classification extends to SO(32) the results of ref.~\cite{Katsuki:1989bf} obtained in the context of
$\text{E}_8\times\text{E}_8'$ heterotic orbifolds. Comparing the numbers of inequivalent SO(32) models to those
presented in ref.~\cite{Katsuki:1989bf}, we find that there are more inequivalent models in the SO(32) heterotic
string for $\mathbb{Z}_N$ orbifolds with $N\le7$ and, conversely, the number of inequivalent models for
orbifolds with $N>7$ is larger in the case of $\text{E}_8\times\text{E}_8'$. This difference becomes important if
Wilson lines are present, since then one can interpret the action of the shift plus the associated Wilson
line(s) locally around every fixed point as a new shift. However, this new shift must be one of the set
of inequivalent shift vectors we have before Wilson lines are switched on\footnote{For more information
 about the concept of fixed point equivalent models, see ref.~\cite{Gmeiner:2002es}.}. In this sense, for $N\le7$ the
SO(32) orbifolds lead to a richer variety of models.

\begin{table}[!h]
\begin{center}
\begin{tabular}{|l|r|r|}
\hline
                   & \multicolumn{2}{c|}{\# ineq. models in}\\
 $\mathbb{Z}_N$    &  SO(32)           & $\text{E}_8\times\text{E}_8'$\\
 \hline
 $\mathbb{Z}_3$    &    6              &    5      \\ 
 $\mathbb{Z}_4$    &   16              &   12      \\
 $\mathbb{Z}_6$-I  &   80              &   58      \\
 $\mathbb{Z}_6$-II &   75              &   61      \\
 $\mathbb{Z}_7$    &   56              &   40      \\
 $\mathbb{Z}_8$-I  &  196              &  246      \\
 $\mathbb{Z}_8$-II &  194              &  248      \\
$\mathbb{Z}_{12}$-I& 2295              & 3026      \\
$\mathbb{Z}_{12}$-II&2223              & 3013      \\
\hline
\end{tabular}
\end{center}
\caption{Comparison between the number of inequivalent $\mathbb{Z}_N$ orbifold models in the SO(32)
heterotic string and in the $\text{E}_8\times\text{E}_8'$ heterotic string~\cite{Katsuki:1989bf}.}
\label{tab:compareSO32andE8xE8}
\end{table}

 \begin{table}[!h]
 \begin{center}
 \begin{tabular}{|l|r|r|r|}
 \hline
                    &\multicolumn{1}{c|}{\# models with} & \multicolumn{2}{c|}{\# models with}\\
  $\mathbb{Z}_N$    & anomalous $\text{U}_1$         & $\bs{16}$ of SO(10) & $\bs{32}$ of SO(12)\\

  \hline
  $\mathbb{Z}_3$    &           5                     &0                  & 0          \\ 
  $\mathbb{Z}_4$    &          12                     &2                  & 0          \\
  $\mathbb{Z}_6$-I  &          76                     &4                  & 4          \\
  $\mathbb{Z}_6$-II &          65                     &10                 & 3          \\
  $\mathbb{Z}_7$    &          55                     &2                  & 0          \\
  $\mathbb{Z}_8$-I  &         193                     &12                 & 0          \\
  $\mathbb{Z}_8$-II &         166                     &11                 & 7          \\
 $\mathbb{Z}_{12}$-I&        2269                     &80                 & 36         \\
 $\mathbb{Z}_{12}$-II&       2097                     &116                & 10         \\
 \hline
 \end{tabular}
 \end{center}
 \caption{Numbers of inequivalent $\mathbb{Z}_N$ orbifold models of the SO(32) heterotic string containing at
   least one spinor of SO(10) or SO(12). Spinors of bigger groups do not appear in orbifold models of the
   SO(32) heterotic string. We also present the number of models having an anomalous $\text{U}_1$ factor as
   part of the gauge group.}
 \label{tab:ineq_models}
 \end{table}

In the second column of table~\ref{tab:ineq_models} we present the number of models having an anomalous
$\text{U}_1$. As explained in section~\ref{subsec:Z4_classification},
all $\text{U}_1$ factors are consistent with the anomaly
conditions,~eq.~(\ref{eq:Anomaly_conditions}). Most of the orbifold models of the SO(32) heterotic string
contain an anomalous $\text{U}_1$.

From the phenomenological point of view, the $\text{SO}(32)$ heterotic string has been considered to be a
less promising starting point than the $\text{E}_8\times\text{E}_8'$ theory, one of the reasons being that
one did not expect spinor representations to be present in the spectrum. As first shown by
ref.~\cite{Choi:2004wn}, it is possible to obtain spinor representations in orbifold models of
the SO(32) heterotic string from the twisted sectors. In the third and fourth columns of table~\ref{tab:ineq_models}, we
list the number of models for each $\mathbb{Z}_N$ orbifold in which there is at least one
$\bs{16}$ spinor of SO(10) or one $\bs{32}$ spinor of SO(12), respectively. 
As we will explain in the next section, the mass formula forbids the appearance of
spinors of SO(14) or bigger groups in orbifold models of the SO(32) heterotic string.

\newpage
\section{Spinors in SO(32) Orbifold Models}
\label{sec:spinors}

In the light of recent developments, $\text{SO}(10)$ GUTs are attractive candidates for a theory beyond
the Standard Model \cite{Georgi:1974my, Fritzsch:1974nn, Nilles:2004ej}. In orbifold constructions, GUTs may be realized in an
intermediate picture
\cite{Forste:2004ie, Kobayashi:2004ud, Kobayashi:2004ya, Buchmuller:2004hv, Buchmuller:2005jr}, keeping
their successful predictions and avoiding the problems, from which GUTs in 4 dimensions generically
suffer. Therefore, we are naturally led to look for orbifold models containing the spinor of $\text{SO}(10)$.
The SO(10) gauge group can then be broken to the Standard Model gauge group by the
inclusion of Wilson lines.

We investigate here the possibility of having the 16-dimensional spinor representation of SO(10).
In the standard basis, the simple roots of SO(10) can be written as

\begin{center}
\begin{tabular}{cc}
$\alpha_1 = \left(1,-1,\phantom{-}0,\phantom{-}0,\phantom{-}0\right)$ & 
$\alpha_4 = \left(0,\phantom{-}0,\phantom{-}0,\phantom{-}1,-1\right)$\\
$\alpha_2 = \left(0,\phantom{-}1,-1,\phantom{-}0,\phantom{-}0\right)$ &
$\alpha_5 = \left(0,\phantom{-}0,\phantom{-}0,\phantom{-}1, \phantom{-}1\right)$\\
$\alpha_3 = \left(0,\phantom{-}0,\phantom{-}1,-1,\phantom{-}0\right)$
\end{tabular}
\end{center}

In this basis, the highest weight of the $\bs{16}$ is given by the 5-dimensional vector
$\left(\tfrac{1}{2}^5\right)$. This vector must be part of a 16-dimensional vector
$p_{sh}$ as follows
\begin{equation}
p_{sh}=p+mV=\left(\tfrac{1}{2} ^5, a_1, a_2,..., a_{11} \right),
\label{eq:gral_p}
\end{equation}
where $p\in\Lambda_{\text{Spin(32)}/\mathbb{Z}_2}$, V is a shift, $m\in\mathbb{N}$ is the number of the
studied sector, and the numbers $a_i$ are selected so that $p_{sh}$ fulfills $N
p_{sh}\in\Lambda_{\text{Spin(32)}/\mathbb{Z}_2}$ and the mass formula for massless states
\begin{equation}
p_{sh}^{2}= 2(1-\tilde{N}-\delta c),
\label{eq:mass}
\end{equation}
where $\delta c$ is the shift of the zero point energy, and $\tilde{N}$ is the number operator, as explained 
in ref.~\cite{Forste:2004ie}. It is
important to notice that there can be more than one combination of different $a_i$'s for which the
resulting $p_{sh}$ fulfills all the conditions.

The first consequence of the form of $p_{sh}$ eq.~(\ref{eq:gral_p}) is that one cannot get the $\bs{16}$ of
SO(10) in the untwisted sector ($m=0$), since it only consists of the roots of SO(32), which
can be expressed by the 480 vectors $(\underline{\pm 1,\,\pm 1,\;0^{14}})$.

As a second consequence, one finds that it is not possible to get the $\bs{16}$ of SO(10) in the
$\mathbb{Z}_3$ orbifold. The first five entries of $3 p_{sh}$ are half-integer and thus, since $3
p_{sh}\in\Lambda_{\text{Spin(32)}/\mathbb{Z}_2}$, the remaining 11 entries must also be half-integer,
i.e. $3a_i\in \mathbb{Z}+\tfrac{1}{2}$. Assuming the smallest value $3a_i=\tfrac{1}{2}$, it follows that
$p_{sh}^2\ge\tfrac{5}{4}+\tfrac{11}{36}=\tfrac{14}{9}$.
In the case of $\tilde{N}=0$, for any twisted sector the right-hand side of eq.~(\ref{eq:mass}) is equal
to $\tfrac{4}{3}<\tfrac{14}{9}$, which forbids the appearance of $p_{sh}$ in the spectrum of any
$\mathbb{Z}_3$ orbifold model. This does not change for $\tilde{N}\neq 0$ since the value of
the right-hand side of eq.~(\ref{eq:mass}) in this case is even smaller.

In particular, one can easily find all shift vectors which produce SO(10) spinors in the first
twisted sector. From eq.~(\ref{eq:gral_p}), for all $\mathbb{Z}_N$ orbifolds with $N>3$ the shift(s)
giving rise to the $\bs{16}$ of SO(10) in the first twisted sector ($m=1$) can be written simply as
\begin{equation}
V=p_{sh}-p \stackrel{p = 0}{\longrightarrow}p_{sh},
\label{eq:shift16}
\end{equation}
where we have chosen $p=0$ because two shifts are equivalent if they differ by an arbitrary
lattice vector. This shift is automatically modular invariant. 

Finding the highest weight of the $\bs{16}$ of SO(10) is a necessary condition for its existence in the
spectrum, but it is not sufficient to guarantee the presence of an SO(10) gauge group. One also needs to
compute the gauge group induced by eq.~(\ref{eq:shift16}), which can be done by simply using the patterns
given in appendix~\ref{sec:ansatz}.

As an example, we consider the $\mathbb{Z}_4$ orbifold. The only possible shift consistent with
$4V\in\Lambda_{\text{Spin(32)}/\mathbb{Z}_2}$ and eqs.~(\ref{eq:mass}) and (\ref{eq:shift16}) with
$\delta c = \tfrac{5}{16}$ is 
\begin{equation}
V=p_{sh}=\left(\left(\tfrac{1}{2}\right)^5,\,\left(\tfrac{1}{4}\right)^2,\,0^{9} \right).
\label{eq:Z4shiftwith16}
\end{equation}
This shift is equivalent to V$^{(5)}$ of table~\ref{tab:Z4_classification} up to lattice vectors and Weyl
reflections. One can also verify that this shift provides indeed several copies of the $\bs{16}$ of SO(10)
in the first twisted sector. 

It is phenomenologically attractive to have $\bs{16}$'s of $\text{SO}(10)$ in the first twisted sector. 
Since the $\bs{16}$-plets of the first twisted sector are localized in all six compact dimensions, the
inclusion of Wilson lines will break the gauge group and will reduce the degeneracy of the fixed points, but it will not 
project out
parts of a $\bs{16}$-plet. Therefore, models with this feature can lead to potentially realistic string models.

An indirect method to obtain the $\bs{16}$ of SO(10) is to switch on Wilson lines in models having spinor representations of bigger
groups, like SO(12). In general, for SO($2n$) groups, the highest weight of the corresponding spinor is a
solution of the eq.~(\ref{eq:mass}) of the form
\begin{equation}
p_{sh}=p+mV=\left(\tfrac{1}{2} ^n, a_1, a_2,..., a_{16-n} \right).
\label{eq:gral_p_D_n}
\end{equation}
By inspecting all possible values that $\delta c$ and $\tilde{N}$ can take in the twisted sectors for all $\mathbb{Z}_N$
orbifolds, one can see that $p_{sh}^2=2(1-\tilde{N}-\delta c)\le \tfrac{31}{18}$ in the mass
equation~(\ref{eq:mass}). This means that the spinor representation of $\text{SO}(2n)$ for $n\ge 7$ given by
eq.~(\ref{eq:gral_p_D_n}) is not allowed, because it is forbidden by the mass equation.

There are indeed some models with the $\bs{32}$ spinor representation of SO(12), as shown in
table~\ref{tab:ineq_models}. One might switch on Wilson lines on them in search of realistic models.

\section{Three-Family Orbifold Models}
\label{subsec:3_family}

\subsection{The $\boldsymbol{\mathbb{Z}_4}$ Orbifold}
\label{subsec:Z4_model}

\begin{figure}[!h]
\begin{center}
\subfigure[$\theta^1$ twisted sector.]{
   \input{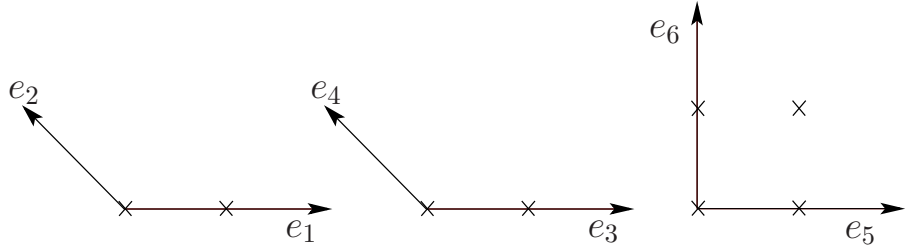}
}
\subfigure[$\theta^2$ twisted sector.]{
   \input{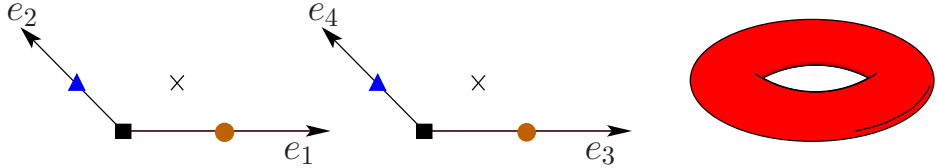}
}
\end{center}
\vspace{-0.5cm}
\caption{Twisted sectors of the $\mathbb{Z}_4$ orbifold.}
\label{fig:Z4_geometry}
\end{figure}

As our illustrative example, we consider the $\mathbb{Z}_4$ orbifold.
One choice for the 6 dimensional lattice \cite{Kobayashi:1991rp} is the SO(5)$^2\times$SO(4) root
lattice as shown in fig. \ref{fig:Z4_geometry}. The point group $\mathbb{Z}_4$ is generated by $\theta$ 
which acts as a simultaneous rotation of 90${}^\circ$ in two of the three 2-tori and a rotation of
180${}^\circ$ in the third one; this corresponds to the twist vector
\begin{equation}
v = \frac{1}{4}\left(1,\,\,\,1,\,\,-2 \right).
\label{eq:twist_Z4}
\end{equation}

{\bf{Fixed point structure.}} On the torus, the action of $\theta^1$ has $2\times2\times4 = 16$ fixed
points, see fig.~\ref{fig:Z4_geometry}(a). The twisted sector corresponding to the action of $\theta^3$ 
gives the anti-particles of the $\theta^1$ sector, so we will not consider it separately. 

The element $\theta^2$ of the point group acts non-trivially only in two of the three complex planes, see 
fig.~\ref{fig:Z4_geometry}(b). Thus, the strings are localized only in 4 of the 6 compact dimensions and are 
free to move in the last torus. 
For convenience, we shall refer to these fixed tori as fixed
points. Of the $4\times 4 = 16$ fixed points of the $\theta^2$ sector, only 
\begin{center}
( ${\scriptstyle \blacksquare}$, ${\scriptstyle \blacksquare}$ ),\;\; ( ${\scriptstyle \blacksquare}$,
\textcolor{brown}{$\bullet$} ),\;\; ( \textcolor{brown}{$\bullet$}, ${\scriptstyle \blacksquare}$ ),\;\;
( \textcolor{brown}{$\bullet$}, \textcolor{brown}{$\bullet$} )
\end{center}
are also invariant under the action of $\theta$. The remaining 12 points are pairwise related by
$\theta$ and therefore form pairs
\begin{center}
( ${\scriptstyle \blacksquare}$, \textcolor{blue}{$\blacktriangle$} ) $\leftrightarrow$ ( ${\scriptstyle \blacksquare}$, $\mathbf{\times}$ ),\;\;
( \textcolor{blue}{$\blacktriangle$}, ${\scriptstyle \blacksquare}$ ) $\leftrightarrow$ ( $\mathbf{\times}$, ${\scriptstyle \blacksquare}$ ),\;\;
( \textcolor{blue}{$\blacktriangle$}, \textcolor{blue}{$\blacktriangle$} ) $\leftrightarrow$ ( $\mathbf{\times}$, $\mathbf{\times}$ ),\\
( \textcolor{blue}{$\blacktriangle$}, \textcolor{brown}{$\bullet$} ) $\leftrightarrow$ ( $\mathbf{\times}$, \textcolor{brown}{$\bullet$} ),\;\;
( \textcolor{blue}{$\blacktriangle$}, $\mathbf{\times}$ ) $\leftrightarrow$ ( $\mathbf{\times}$, \textcolor{blue}{$\blacktriangle$} ),\;\;
( \textcolor{brown}{$\bullet$}, \textcolor{blue}{$\blacktriangle$} ) $\leftrightarrow$ ( \textcolor{brown}{$\bullet$}, $\mathbf{\times}$ ).
\end{center}
In this way, these 12 fixed points of the $\theta^2$ sector collapse to 6 by the action of
the orbifold. This leaves an effective number of $4 + 6 = 10$ fixed points in the second twisted
sector.

{\bf{Wilson lines.}} Of the 16 $\mathbb{Z}_4$ models, none has 3 families of quarks and leptons. In order
to reduce the number of families and to further break the gauge symmetry, we need Wilson lines
\cite{Ibanez:1986tp}. The number and the order of Wilson lines one can add in a specific orbifold model
is dictated by the geometry of the underlying compactification. In our case, we can have only 4 Wilson
lines $A_1$, $A_3$, $A_5$, and $A_6$  of order 2 corresponding to the directions $e_1$, $e_3$, $e_5$, and
$e_6$, respectively.

\begin{figure}[!h]
\center\input{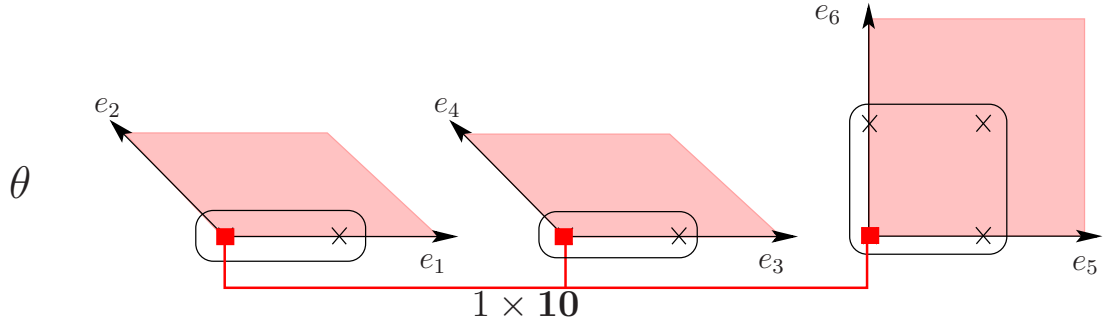}
\caption{Localization of the three generations with SU(5) gauge group. There are two families in the bulk
and one family localized in the origin. The boxes correspond to the degeneracy of the fixed points
without Wilson lines. This degeneracy has been lifted by the Wilson lines in the $e_1$, $e_3$, $e_5$ and $e_6$
directions.}
\label{fig:Z4_3gen_model}
\end{figure}
\vskip 1cm
{\bf{Three-generation model.}} As a toy model, we present a 3-generation SU(5) model. When considering
the models presented in table~\ref{tab:Z4_classification}, the shift vector
V$^{(14)}$ seems quite promising. This model has an SU(5) gauge symmetry and, most importantly, the
localization of the generations gives some clues. There are two $\bs{10}$'s in the bulk and sixteen $\bs{10}$'s attached to
the fixed points of the first twisted sector. We focus our attention on the $\bs{10}$ because
the representation $\bsb{5}$ generically come with the same multiplicity due to anomaly cancellation in orbifold
models. By a clever choice of the four Wilson lines, the degeneracy of the fixed points can be lifted, so
that the number of families is reduced from 16 in the twisted sectors to 1, and both families of the
untwisted sector survive, as depicted in fig.~\ref{fig:Z4_3gen_model}:
\begin{center}
\begin{tabular}{l}
$A_1=\left(\phantom{-}\frac{5}{2}^2,\phantom{-}0^5,-\2 ^2 -1^2,\phantom{-}3,-1^4  \right)$, \\
$A_3=\left(-3^2,\phantom{-}0^5,-3^2,-2,-3,-\frac{5}{2},\phantom{-}\frac{3}{2},-\frac{5}{2},-2,\phantom{-}\frac{1}{2}\right)$,\\
$A_5=\left(\phantom{-}\2 ^2,\phantom{-}0^5,\phantom{-}\2 ^2,\phantom{-}2,\phantom{-}\frac{3}{2}^3,\phantom{-}2,\phantom{-}\2,\phantom{-}2 \right)$,\\
$A_6=\left(\phantom{-}3,\phantom{-}\frac{7}{2},\phantom{-}0^5,-1,-\frac{5}{2},-2,-\frac{5}{2}^6
\right)$.
\end{tabular}
\end{center}

The combined action of the shift and the Wilson lines leads to the gauge group
$\text{SU}(5)\times\text{SU}(2)^5\times\text{U}(1)^7$, where the first U(1) is anomalous. The complete spectrum of this
model is given in table \ref{tab:spectrum_of_SU5_model_in_Z4}. The main objective of the present publication being the 
clarification of some outstanding issues in the heterotic $\text{SO}(32)$ theory, we will not explore the phenomenology of this model in detail.

\begin{table}[!h]
\begin{center}
\begin{tabular}{cc|c|c|c|cc}
&U&$\theta^1$&$\theta^2$&sum&states&\\
\hline
& 2 & 1 &   & 3 & $(\bs{10};\;\bs{1},\;\bs{1},\;\bs{1},\;\bs{1},\;\bs{1})$&\\
& 3 & 5 &   & 8 & $(\phantom{\bs{1}}\bsb{5};\;\bs{1},\;\bs{1},\;\bs{1},\;\bs{1},\;\bs{1})$&\\
& 1 &   & 4 & 5 & $(\bs{\phantom{1}5};\;\bs{1},\;\bs{1},\;\bs{1},\;\bs{1},\;\bs{1})$&\\
& 10& 37& 4 & 51& $(\bs{\phantom{1}1};\;\bs{1},\;\bs{1},\;\bs{1},\;\bs{1},\;\bs{1})$&\\
\hline
&& 12& 4 & 16& $(\bs{\phantom{1}1};\;\bs{2},\;\bs{1},\;\bs{1},\;\bs{1},\;\bs{1})$&\\
&& 12& 4 & 16& $(\bs{\phantom{1}1};\;\bs{1},\;\bs{2},\;\bs{1},\;\bs{1},\;\bs{1})$&\\
&& 12& 4 & 16& $(\bs{\phantom{1}1};\;\bs{1},\;\bs{1},\;\bs{2},\;\bs{1},\;\bs{1})$&\\
&& 12& 4 & 16& $(\bs{\phantom{1}1};\;\bs{1},\;\bs{1},\;\bs{1},\;\bs{2},\;\bs{1})$&\\
&& 12& 4 & 16& $(\bs{\phantom{1}1};\;\bs{1},\;\bs{1},\;\bs{1},\;\bs{1},\;\bs{2})$&\\
\end{tabular}
\end{center}  
\vspace{-0.5cm}
\caption{The spectrum of a $\mathbb{Z}_4$ toy model with 3 generations of SU(5).}
\label{tab:spectrum_of_SU5_model_in_Z4}
\end{table}

We would like to stress three features of this model. To the best of our knowledge,
this is the first three-generation model in the context of the SO(32) orbifold published in
the literature. The shift V$^{(14)}$ that we used for this three-family model does not appear in
ref.~\cite{Choi:2004wn}. This model shows clearly the possibility to compute promising models
through orbifolds of the SO(32) heterotic string.

\subsection{Model in the $\boldsymbol{\mathbb{Z}_6}$-II Orbifold}
\label{subsec:Z6II_model}
\begin{figure}[!h]
\center\input{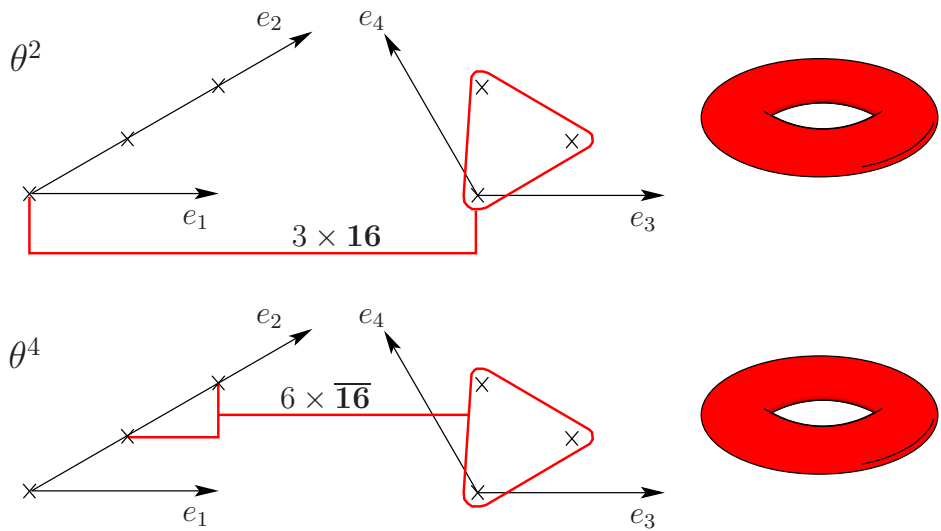}
\vspace{-1cm}
\caption{Localization of the 3 generations with SO(10) gauge group. There are 3 families in the
  second twisted sector and 6 anti-families in the fourth twisted sector, giving a net number of 3
  (anti-)families free to move in two of the six compactified dimensions. The box corresponds to the
degeneracy of the fixed points in the SU(3) 2-torus.}
\label{fig:Z6_II_3gen_model}
\end{figure}

In the $\mathbb{Z}_6$-II orbifold, one possible choice of the 6 dimensional lattice is
G$_2\times$SU(3)$\times$SO(4). For further details on the geometry and the fixed point structure, see ref.~\cite{Kobayashi:2004ya}. Even without the inclusion of Wilson lines, 
we find toy models with 3 generations. For instance, using
\begin{center}
$V^{(30)} =
\left(\;\frac{1}{2}^{2},\;-\frac{1}{6}^{5},\;-\frac{1}{3}^{6},\;-\frac{1}{2}^{3}\;\right)$
\end{center}
of those $\mathbb{Z}_6$-II shifts listed on our web page~\cite{SO32:webseite}, we obtain a model with 
3 generations of SO(10). Their localization is illustrated in fig.~\ref{fig:Z6_II_3gen_model}.

The families are localized as follows: there are three $\bs{16}$'s of
SO(10) in the second twisted sector, whereas there are six $\bsb{16}$'s in the fourth twisted
sector. Since the families are located in the second
and fourth twisted sectors, where two of the six compactified dimensions are left invariant by the
orbifold action, the families are free to move in six dimensions.

Even though this model is not realistic, it illustrates how easily one can obtain orbifold models with three 
families in the context of the SO(32) heterotic string. Therefore, using Wilson lines, potentially realistic models may be derived. 

\section{Conclusions and Outlook}

As we have seen, model building with the heterotic SO(32) theory might
be as exciting as that with its more famous brother: the $\text{E}_8\times\text{E}_8'$
string, see e.g.~\cite{Faraggi:1989ka,Kobayashi:2004ud,Kobayashi:2004ya,Buchmuller:2004hv,Braun:2005ux,Buchmuller:2005jr,Bouchard:2005ag,Blumenhagen:2006ux}. It opens new roads for explicit constructions that
should be explored as a vital part of the string landscape.

We were somewhat surprised about the frequency of the appearance of
spinor representations of SO(2$n$) gauge groups. These spinors might
be an important tool to implement the family structure of
SU(3) $\times$ SU(2) $\times$ U(1) models. In addition they are an important
ingredient for a possible understanding of the SO(32) 
heterotic type I duality in $d=4$ space time dimensions. We know that
these spinors do not appear in the perturbative type~I theory. Thus the
mentioned duality will need the implementation of nonperturbative
effects. 

Our classification of the $\mathbb{Z}_N$ orbifolds of the SO(32) theory 
completes a basic building block for further model constructions.
We understand this as a contribution to the study of the string
landscape in the spirit of the ``String Vacuum Project''~\cite{SVP}. A 
further step in this program would be the implementation of 
Wilson lines that leads to enormous complexity and a huge number
of models (comparable to that of the $\text{E}_8\times\text{E}_8'$ string).
An exploration of this large region of the landscape is currently
beyond our capabilities. We therefore gather our present results 
and make them available to the public on our web page~\cite{SO32:webseite}, such 
that interested people could share our knowledge and contribute to the enterprise.

\bigskip

\noindent {\bf Acknowledgments} 
 
\noindent 
It is a pleasure to thank K.~S.~Choi, S.~F\"orste and M.~Ratz for valuable discussions. This work was partially
supported by the European Union 6th Framework Program\\
\mbox{MRTN-CT-2004-503369} ``Quest for Unification'' and \mbox{MRTN-CT-2004-005104}
``ForcesUniverse''. A.~W. was supported in part by DOE grant DOE/ER/01545-866.

\newpage
\appendix
\def\theequation{\thesection.\arabic{equation}}
\setcounter{equation}{0}

\begin{landscape}
\setlength{\voffset}{1.6cm}
\setlength{\textheight}{24.5cm}
\setlength{\tabcolsep}{2pt}
\setlength{\LTcapwidth}{\textheight}
\section{Table of $\mathbb{Z}_4$ orbifold models}
\label{sec:Z4_models}

{\small

\begin{longtable}{|c|c|p{2.85cm}|c|p{4.3cm}|p{3.7cm}|p{4.2cm}p{0.5mm}|}
\multicolumn{8}{c}{\label{tab:Z4_classification}}\\
\caption{All admissible models for the $\mathbb{Z}_4$ orbifold of the SO(32)
  heterotic string without Wilson lines. For each model, we list the shift vector, its classification according to
  ref.~\cite{Choi:2004wn} and the matter content displayed in sectors.}\\
\hline
 & &\centering{\small{Ref.~\cite{Choi:2004wn}}} & & & \multicolumn{2}{c}{Twisted  matter} &\\
\cline{6-8}
\# & Shift &\centering{\small{classification}} & 4D gauge group & \centering{Untwisted matter} &
\centering{$T_1$} & \centering{$T_2$} &\\
\hline
\hline
\endfirsthead
\hline
 & &\centering{\small{Ref.~\cite{Choi:2004wn}}} & & & \multicolumn{2}{c}{Twisted  matter} &\\
\cline{6-8}
\# & Shift &\centering{\small{classification}} & 4D gauge group & \centering{Untwisted matter} &
\centering{$T_1$} & \centering{$T_2$} &\\
\hline
\hline
\endhead
\multicolumn{8}{r}{\footnotesize{\emph{Continued...}}}
\endfoot
\endlastfoot

1 & $\left( 0^2, \text{-}\fr{1}{2}, \text{-}\fr{3}{4}^2, 1, 0^{10} \right)$ & 
\centering{$n_1 = 2, n_2 = 1$\newline \small{Vectorial shift} } &  SO$_{26} \times$ SU$_2 \times$ U$_{1A}\times$U$_1$ &
\centering{
$1(\bs{26},\bs{1})_{\text{-}\2,\text{-}\2} + 1(\bs{26},\bs{1})_{\2,\2} +
2(\bs{26},\bs{2})_{\text{-}\2,\4} + 2(\bs{1},\bs{2})_{0,\text{-}\fr{3}{4}} +
2(\bs{1},\bs{2})_{1,\4} + 1(\bs{1},\bs{1})_{1,\text{-}\2} +
1(\bs{1},\bs{1})_{\text{-}1,\2} $  } & \centering{
 $16(\bs{26},\bs{1})_{\text{-}\2,\text{-}\fr{1}{8} } +
32(\bs{1},\bs{2})_{0,\text{-}\fr{3}{8} } + 16(\bs{1},\bs{1})_{1,\text{-}\fr{1}{8} }
+ 80(\bs{1},\bs{1})_{0,\fr{3}{8} }$ } & \centering{
 $10(\bs{26},\bs{1})_{\text{-}\2,\4} + 6(\bs{26},\bs{1})_{\2,\text{-}\4} +
32(\bs{1},\bs{2})_{0,0} + 10(\bs{1},\bs{1})_{0,\text{-}\fr{3}{4}} +
10(\bs{1},\bs{1})_{1,\4} + 6(\bs{1},\bs{1})_{\text{-}1,\text{-}\4} +
6(\bs{1},\bs{1})_{0,\fr{3}{4}}$ }& \\
\hline
2 & $\left( 0^2, \text{-}\fr{1}{2}^2, \fr{1}{2}, \fr{1}{4}, \text{-}\fr{3}{4}, 1, 0^{8} \right)$ & 
\centering{$n_1 = 2, n_2 = 3$ \newline \small{Vectorial shift} }&  SO$_{22} \times$ SU$_4 \times$ SU$_2 \times$ U$_1$ &
\centering{
$1(\bs{22},\bs{6},\bs{1})_{0} + 2(\bs{22},\bs{1},\bs{2})_{\4} +
2(\bs{1},\bs{6},\bs{2})_{\text{-}\4} + 1(\bs{1},\bs{1},\bs{1})_{\text{-}\2} +
1(\bs{1},\bs{1},\bs{1})_{\2} $  } & \centering{
 $16(\bs{1},\bsb{4},\bs{2})_{\text{-}\8 } +
32(\bs{1},\bs{4},\bs{1})_{\8} $ } & \centering{
 $10(\bs{22},\bs{1},\bs{1})_{\text{-}\4} + 6(\bs{22},\bs{1},\bs{1})_{\4} +
10(\bs{1},\bs{6},\bs{1})_{\4} + 6(\bs{1},\bs{6},\bs{1})_{\text{-}\4} +
32(\bs{1},\bs{1},\bs{2})_{0}$ }& \\
\hline
3 & $\left(0^{2},\text{-}\frac{3}{4}^2,\frac{1}{4}^{3},\frac{9}{4},\text{-}2,0^{7}\right)$
 & 
\centering{$n_1 = 6, n_2 = 0$ \newline \small{Vectorial shift}} &  SO$_{20} \times$ SU$_6 \times$ U$_{1A}$ &
\centering{
$2(\bs{20},\bsb{6})_{\text{-}\2} + 1(\bs{1},\bs{15})_{1} +
1(\bs{1},\bsb{15})_{\text{-}1} $  } & \centering{
 $16(\bs{1},\bs{15})_{\4 } +
 80(\bs{1},\bs{1})_{\text{-}\fr{3}{4} }$ } & \centering{$10(\bs{1},\bs{15})_{\text{-}\2} + 6(\bs{1},\bsb{15})_{\2} +
10(\bs{1},\bs{1})_{\fr{3}{2}} + 6(\bs{1},\bs{1})_{\text{-}\fr{3}{2}}$ }& \\
\hline
4 &
$\left(0^{2},\text{-}\frac{1}{4},\text{-}\frac{3}{4},\frac{1}{4}^{3},\text{-}\frac{3}{4},1,0^{7}\right)$ & 
\centering{\small{not classified}\newline \small{Vectorial shift}} &  SO$_{20} \times$ SU$_6 \times$ U$_{1A}$ &
\centering{
$2(\bs{20},\bs{6})_{\text{-}\2} + 1(\bs{1},\bs{15})_{\text{-}1} +
1(\bs{1},\bsb{15})_{1} $  } & \centering{
 $16(\bs{20},\bs{1})_{\text{-}\fr{3}{4} } +
 32(\bs{1},\bsb{6})_{\text{-}\fr{1}{4} }$ } & \centering{$10(\bs{1},\bsb{15})_{\text{-}\2} + 6(\bs{1},\bs{15})_{\2} +
10(\bs{1},\bs{1})_{\fr{3}{2}} + 6(\bs{1},\bs{1})_{\text{-}\fr{3}{2}}$ }& \\
\hline
5 & $\left( 0^2, \text{-}\fr{1}{2}^2, \fr{1}{2}^3, \fr{1}{4}, \fr{9}{4}, \text{-}2, 0^{6} \right)$ & 
\centering{$n_1 = 2, n_2 = 5$ \newline \small{Vectorial shift}}&  SO$_{18} \times$ SO$_{10} \times$ SU$_2 \times$ U$_{1A}$ &
\centering{
$1(\bs{18},\bs{10},\bs{1})_{0} + 2(\bs{18},\bs{1},\bs{2})_{\text{-}\2} +
2(\bs{1},\bs{10},\bs{2})_{\2} + 1(\bs{1},\bs{1},\bs{1})_{1} +
1(\bs{1},\bs{1},\bs{1})_{\text{-}1} $  } & \centering{
 $16(\bs{1},\bs{16},\bs{1})_{\text{-}\fr{1}{4} }$ } & \centering{
 $10(\bs{18},\bs{1},\bs{1})_{\text{-}\2} + 6(\bs{18},\bs{1},\bs{1})_{\2} +
10(\bs{1},\bs{10},\bs{1})_{\2} + 6(\bs{1},\bs{10},\bs{1})_{\text{-}\2} +
32(\bs{1},\bs{1},\bs{2})_{0}$ }& \\
\hline
6 & $\left( 0^2, \text{-}\fr{1}{2}^2, \fr{1}{4}^5, \text{-}\fr{3}{4}, 1, 0^{5} \right)$ & 
\centering{$n_1 = 6, n_2 = 2$\newline \small{Vectorial shift}} &  SO$_{16} \times$ SU$_2^2
\times$ SU$_6 \times$ U$_{1A}$ &
\centering{
$1(\bs{16},\bs{2},\bs{2},\bs{1})_{0} + 2(\bs{16},\bs{1},\bs{1},\bsb{6})_{\text{-}\2} +
1(\bs{1},\bs{1},\bs{1},\bs{15})_{1} + 1(\bs{1},\bs{1},\bs{1},\bsb{15})_{\text{-}1} +
2(\bs{1},\bs{2},\bs{2},\bs{6})_{\2} $  } & \centering{
 $16(\bs{1},\bs{1},\bs{2},\bs{6})_{\text{-}\fr{1}{4} } + 32(\bs{1},\bs{2},\bs{1},\bs{1})_{\text{-}\fr{3}{4} }$ } & \centering{
 $10(\bs{1},\bs{1},\bs{1},\bsb{15})_{\2} + 6(\bs{1},\bs{1},\bs{1},\bs{15})_{\text{-}\2} +
10(\bs{1},\bs{1},\bs{1},\bs{1})_{\text{-}\fr{3}{2}} + 6(\bs{1},\bs{1},\bs{1},\bs{1})_{\fr{3}{2}}$ }& \\
\hline
7 & $\left( 0^2, \text{-}\fr{1}{2}^4, \fr{1}{2}^3, \fr{1}{4}, \fr{5}{4}, \text{-}1, 0^{4} \right)$ & 
\centering{$n_1 = 2, n_2 = 7$\newline \small{Vectorial shift}} &  SO$_{14} \times$ SO$_{14} \times$ SU$_2 \times$ U$_1$ &
\centering{
$1(\bs{14},\bs{14},\bs{1})_{0} + 2(\bs{14},\bs{1},\bs{2})_{\4} +
2(\bs{1},\bs{14},\bs{2})_{\text{-}\4} + 1(\bs{1},\bs{1},\bs{1})_{\text{-}\2} +
1(\bs{1},\bs{1},\bs{1})_{\2} $  } & &  \centering{
 $10(\bs{14},\bs{1},\bs{1})_{\text{-}\4} + 6(\bs{14},\bs{1},\bs{1})_{\4} +
10(\bs{1},\bs{14},\bs{1})_{\4} + 6(\bs{1},\bs{14},\bs{1})_{\text{-}\4} +
32(\bs{1},\bs{1},\bs{2})_{0}$ }& \\
\hline
8 & $\left( 0^2, \text{-}\fr{1}{2}^4, \fr{1}{4}^5, \fr{9}{4}, \text{-}2, 0^{3} \right)$ & 
\centering{$n_1 = 6, n_2 = 4$ \newline \small{Vectorial shift}}&  SO$_{12} \times$ SO$_{8} \times$ SU$_6 \times$ U$_{1A}$ &
\centering{
$1(\bs{12},\bs{8},\bs{1})_{0} + 2(\bs{12},\bs{1},\bsb{6})_{\text{-}\2} +
2(\bs{1},\bs{8},\bs{6})_{\2} + 1(\bs{1},\bs{1},\bs{15})_{1} +
1(\bs{1},\bs{1},\bsb{15})_{\text{-}1} $  } & \centering{
 $16(\bs{1},\bs{8},\bs{1})_{\text{-}\fr{3}{4} }$ } & \centering{
 $10(\bs{1},\bs{1},\bs{15})_{\text{-}\2} + 6(\bs{1},\bs{1},\bsb{15})_{\2} +
10(\bs{1},\bs{1},\bs{1})_{\fr{3}{2}} + 6(\bs{1},\bs{1},\bs{1})_{\text{-}\fr{3}{2}}$ }& \\
\hline
9 & $\left( 0^2, \text{-}\fr{1}{2}, \text{-}\fr{3}{4}, \fr{1}{4}^8, \fr{9}{4},\text{-}2,0^2 \right)$ & 
\centering{$n_1 = 10, n_2 = 1$ \newline \small{Vectorial shift}} &  SO$_{10} \times$ SU$_{10} \times$ U$_{1A}\times$U$_1$ &
\centering{
$2(\bs{10},\bsb{10})_{\text{-}\2,\4} + 1(\bs{10},\bs{1})_{\text{-}1,\text{-}\fr{5}{4}} +
1(\bs{10},\bs{1})_{1,\fr{5}{4}} + 2(\bs{1},\bs{10})_{\text{-}\2,\text{-}\fr{3}{2}} +
2(\bs{1},\bs{10})_{\fr{3}{2},1} + 1(\bs{1},\bs{45})_{1,\text{-}\2} +
1(\bs{1},\bsb{45})_{\text{-}1,\2}$} & \centering{
 $16(\bs{1},\bs{10})_{\text{-}\fr{5}{4},\text{-}\4 } +
32(\bs{1},\bs{1})_{\text{-}\fr{3}{4},\fr{5}{4} }$ } & \centering{$10(\bs{16},\bs{1})_{\text{-}\2,\text{-}\fr{5}{8}} + 6(\bsb{16},\bs{1})_{\2,\fr{5}{8}}$ }& \\
\hline
10 & $\left( 0^2, \text{-}\fr{1}{2}^2, \fr{1}{2}, \fr{1}{4}^9, \fr{9}{4}, 2 \right)$ & 
\centering{$n_1 = 10, n_2 = 3$ \newline \small{Vectorial shift}}&  SU$_{4} \times$ SU$_4 \times$ SU$_{10} \times$ U$_1$ &
\centering{$1(\bs{6},\bs{6},\bs{1})_{0} + 2(\bs{6},\bs{1},\bsb{10})_{\4} +
2(\bs{1},\bs{6},\bs{10})_{\text{-}\4} + 1(\bs{1},\bs{1},\bs{45})_{\text{-}\2} +
1(\bs{1},\bs{1},\bsb{45})_{\2} $  } & \centering{
 $16(\bsb{4},\bs{1},\bs{1})_{\text{-}\fr{5}{8} } +
16(\bs{1},\bs{4},\bs{1})_{\fr{5}{8}} $ } & \centering{
 $10(\bs{4},\bsb{4},\bs{1})_{0} + 6(\bsb{4},\bs{4},\bs{1})_{0}$ }& \\
\hline
11& $\left( \fr{1}{2}^2, \text{-}\fr{1}{4}^{12}, \fr{3}{4}^2 \right)$ & 
\centering{$n_1 = 14, n_2 = 0$ \newline \small{Vectorial shift}}&  SU$_{2} \times$ SU$_2 \times$ SU$_{14} \times$ U$_1$ &
\centering{
$2(\bs{2},\bs{2},\bsb{14})_{\4} + 1(\bs{1},\bs{1},\bs{91})_{\text{-}\2} +
1(\bs{1},\bs{1},\bsb{91})_{\2} $  } & \centering{
 $16(\bs{2},\bs{1},\bs{1})_{\text{-}\fr{7}{8} } +
32(\bs{1},\bs{1},\bs{1})_{\fr{7}{8}} $ } & \centering{
 $10(\bs{1},\bs{2},\bs{14})_{\text{-}\4} + 32(\bs{2},\bs{1},\bs{1})_{0} +
6(\bs{1},\bs{2},\bsb{14})_{\4}$ }& \\
\hline
12& $\left( \fr{1}{2}^2, \fr{1}{4}, \text{-}\fr{1}{4}^{11}, \fr{3}{4}^2 \right)$ & 
\centering{\small{not classified} \newline \small{Vectorial shift}}&  SU$_{2} \times$ SU$_2 \times$ SU$_{14} \times$ U$_{1A}$ &
\centering{
$2(\bs{2},\bs{2},\bs{14})_{\text{-}\2} + 1(\bs{1},\bs{1},\bs{91})_{\text{-}1} +
1(\bs{1},\bs{1},\bsb{91})_{1} $  } & \centering{
 $16(\bs{2},\bs{1},\bs{1})_{\fr{7}{4} } +
16(\bs{1},\bs{1},\bsb{14})_{\text{-}\fr{5}{4}} $ } & \centering{
 $10(\bs{2},\bs{1},\bs{14})_{\text{-}\2} + 32(\bs{1},\bs{2},\bs{1})_{0} +
6(\bs{2},\bs{1},\bsb{14})_{\2}$ }& \\
\hline
13 & $\left( \text{-}\fr{1}{8}, \text{-}\fr{7}{8}, \text{-}\fr{5}{8}, \fr{1}{8}^{11}, \fr{17}{8}^2 \right)$ &
\centering{\small{Spinorial shift}} &  SU$_{15} \times$ U$_{1A}\times$U$_1$ &
\centering{$2\times\bs{105}_{\text{-}5,\fr{3}{2}} + \newline 2\times\bsb{15}_{7,\fr{11}{2}} +
\bs{15}_{2,7} + \bsb{15}_{\text{-}2,\text{-}7} $  } &
\centering{$16\times\bsb{15}_{\text{-}5,\text{-}\fr{13}{4} } \newline +
 80\times\bs{1}_{\text{-}3,\fr{15}{4} }$ }
 & \centering{ $10\times\bsb{15}_{\text{-}8,\fr{1}{2}} + \newline
   10\times\bs{1}_{6,\text{-}\fr{15}{2}}
+ \newline 6\times\bs{15}_{8,\text{-}\fr{1}{2}} + 6\times\bs{1}_{\text{-}6,\fr{15}{2}}$ }& \\
\hline
14 & $\left( \text{-}\fr{9}{8}, \fr{1}{8}, \text{-}\fr{13}{8}, \text{-}\fr{5}{8}^4, \text{-}\fr{7}{8}^9 \right)$ & 
\centering{\small{Spinorial shift} } &  SU$_{11} \times$ SU$_5 \times$ U$_{1A}\times$U$_1$ &
\centering{
$2(\bs{55},\bs{1})_{\text{-}3,\text{-}\fr{5}{2}} + 2(\bsb{11},\bsb{5})_{1,\fr{19}{2}} +
2(\bs{1},\bs{10})_{1,\text{-}\fr{33}{2}} + 1(\bs{11},\bsb{5})_{\text{-}2,7} +
1(\bsb{11},\bs{5})_{2,\text{-}7} $  } & \centering{
 $16(\bs{1},\bs{10})_{\text{-}2,\text{-}\fr{11}{4} } +
32(\bs{1},\bs{1})_{\text{-}3,\fr{55}{4} }$ } & \centering{
 $10(\bsb{11},\bs{1})_{\text{-}2,\text{-}\fr{25}{2}} + 10(\bs{1},\bs{5})_{4,\fr{11}{2}} +
6(\bs{11},\bs{1})_{2,\fr{25}{2}} + 6(\bs{1},\bsb{5})_{\text{-}4,\text{-}\fr{11}{2}}$ }& \\
\hline
15 & $\left( \text{-}\fr{1}{8}, \fr{1}{8}, \text{-}\fr{5}{8}^4, \fr{3}{8}^{5}, \fr{1}{8}^5 \right)$ & 
\centering{\small{Spinorial shift}} &  SU$_{7} \times$ SU$_9 \times$ U$_{1A}\times$U$_1$ &
\centering{$2(\bs{21},\bs{1})_{3,\text{-}\fr{3}{2}} + 2(\bsb{7},\bsb{9})_{\text{-}1,\fr{5}{2}} +
2(\bs{1},\bs{36})_{\text{-}1,\text{-}\fr{7}{2}} + 1(\bs{7},\bsb{9})_{2,1} +
1(\bsb{7},\bs{9})_{\text{-}2,\text{-}1}$ } & \centering{
 $16(\bs{7},\bs{1})_{\text{-}3,\text{-}\fr{3}{4} } +
16(\bs{1},\bs{1})_{3,\fr{21}{4} }$ } & \centering{
 $10(\bsb{7},\bs{1})_{0,\text{-}\fr{9}{2}} + 10(\bs{1},\bs{9})_{\text{-}2,\fr{7}{2}} +
6(\bs{7},\bs{1})_{0,\fr{9}{2}} + 6(\bs{1},\bsb{9})_{2,\text{-}\fr{7}{2}}$ }& \\
\hline
16 & $\left(\fr{3}{8}, \fr{5}{8}, \text{-}\fr{1}{8}^{12}, \fr{15}{8}, \text{-}\fr{3}{8} \right)$ & 
\centering{\small{Spinorial shift}} &  SU$_{3} \times$ SU$_{13} \times$ U$_{1A}\times$U$_1$ &
\centering{$2(\bsb{3},\bs{1})_{3,\text{-}\fr{13}{2}} + 2(\bsb{3},\bsb{13})_{\text{-}1,\fr{11}{2}} +
2(\bs{1},\bs{78})_{\text{-}1,\text{-}\fr{9}{2}} + 1(\bs{3},\bsb{13})_{2,\text{-}1} +
1(\bsb{3},\bs{13})_{\text{-}2,1}$ } & \centering{
 $16(\bs{1},\bs{1})_{2,\text{-}\fr{39}{4} } +
16(\bs{1},\bsb{13})_{\text{-}2,\fr{9}{4} } +
32(\bs{3},\bs{1})_{\text{-}1,\text{-}\fr{13}{4} }$ } & \centering{
 $10(\bsb{3},\bs{1})_{\text{-}2,\text{-}\fr{13}{2}} + 10(\bs{1},\bs{13})_{0,\fr{15}{2}} +
6(\bs{3},\bs{1})_{2,\fr{13}{2}} + 6(\bs{1},\bsb{13})_{0,\text{-}\fr{15}{2}}$ }&\\
\hline
\end{longtable}

}

\end{landscape}

\newpage
\def\theequation{\thesection.\arabic{equation}}
\setcounter{equation}{0}

\section{The General Form of a Shift in $\boldsymbol{\mathbb{Z}_N}$ Orbifolds of the SO(32) Heterotic String}
\label{sec:ansatz}

To obtain the general form of a shift, we use the fact that two shifts are equivalent if they are related
by lattice vectors or by Weyl reflections, i.e. by any permutation of the entries and pairwise sign flips.

In $\mathbb{Z}_N$ orbifolds with {\bf{even $\boldsymbol{N}$}}, one can prove that the most general form of a {\bf{vectorial shift}} is given by
\begin{equation}
V = \frac{1}{N}\left( \left(\pm k\right)^{\alpha}, -(N-k)^{\beta}, 0^{n_0}, 1^{n_1}, \ldots ,
  (N-k)^{n_{(N-k)}-\alpha-\beta},\ldots, \left(\frac{N}{2}\right)^{n_{\left(\frac{N}{2}\right)}} \right),
\end{equation}
where $\alpha, \beta, n_i \in \mathbb{N}$, $\alpha+\beta\in \{0,1\}$,
$k\in\{\frac{N}{2}+1,\frac{N}{2}+2,\ldots,N\}$ and $\sum n_i =16$. It leads to a symmetry breaking in
four dimensions of the general form
\begin{equation}
\textrm{SO(32)} \longrightarrow \textrm{SO(2}n_0\textrm{)}\times \textrm{U(}n_1\textrm{)}\times
\ldots\times\textrm{U}\left(n_{\left(\frac{N}{2}-1\right)}\right)\times \textrm{SO}\left(2n_{\left(\frac{N}{2}\right)}\right).
\end{equation}
On the other hand, for {\bf{even $\boldsymbol{N}$}} the {\bf{spinorial shifts}} can be written in the standard form
\begin{equation}
V = \frac{1}{2N}\left( \left(\pm k\right)^{\alpha}, -(2N-k)^{\beta}, 1^{n_1}, 3^{n_3}, \ldots ,
  (2N-k)^{n_{\left(2N-k\right)}-\alpha-\beta},\ldots, (N-1)^{n_{\left(N-1\right)}} \right),
\end{equation}
with $k\in\{N+1,N+3,\ldots,2N-1\}$, which give rise to the gauge group
\begin{equation}
\textrm{SO(32)} \longrightarrow \textrm{U(}n_1\textrm{)}\times \textrm{U(}n_3\textrm{)}\times
\ldots\times\textrm{U}\left(n_{\left(N-3\right)}\right)\times \textrm{U}\left(n_{\left(N-1\right)}\right).
\end{equation}
For {\bf{$\boldsymbol{N}$ odd}}, the general form of both a vectorial shift and a spinorial one change slightly. As
explained in ref.~\cite{Choi:2004wn}, in this case it is enough to determine either the vectorial or the
spinorial shifts, since one spinorial shift can always be transformed into a vectorial one by the action
of Weyl reflections and lattice vectors. Therefore any shift can be written in general as
\begin{equation}
V = \frac{1}{N}\left( \left(\pm k\right)^{\alpha}, -(N-k)^{\beta}, 0^{n_0}, 1^{n_1}, \ldots ,
  (N-k)^{n_{\left(N-k\right)}-\alpha-\beta},\ldots, \left(\frac{N-1}{2}\right)^{n_{\left(\frac{N-1}{2}\right)}} \right),
\end{equation}
where $k\in\{\frac{N+1}{2},\frac{N+3}{2},\ldots,N\}$. The resulting four dimensional gauge group is
\begin{equation}
\textrm{SO(32)} \longrightarrow \textrm{SO(2}n_0\textrm{)}\times \textrm{U(}n_1\textrm{)}\times
\ldots\times\textrm{U}\left(n_{\left(\frac{N-3}{2}\right)}\right)\times \textrm{U}\left(n_{\left(\frac{N-1}{2}\right)}\right).
\end{equation}


\newpage
\begin{thebibliography}{42}

\bibitem{Polchinski:1995df}
  J.~Polchinski and E.~Witten,
  Nucl.\ Phys.\ B {\bf 460} (1996) 525
  [arXiv:hep-th/9510169].

\bibitem{Dixon:1985jw}
  L.~J.~Dixon, J.~A.~Harvey, C.~Vafa and E.~Witten,
  Nucl.\ Phys.\ B {\bf 261} (1985) 678.

\bibitem{Giedt:2003an}
  J.~Giedt,
  Nucl.\ Phys.\ B {\bf 671} (2003) 133
  [arXiv:hep-th/0301232].

\bibitem{Choi:2004wn}
K.~S.~Choi, S.~Groot Nibbelink and M.~Trapletti,
JHEP {\bf 0412}, 063 (2004)
[arXiv:hep-th/0410232].

\bibitem{Forste:2004ie}
  S.~Forste, H.~P.~Nilles, P.~K.~S.~Vaudrevange and A.~Wingerter,
  Phys.\ Rev.\ D {\bf 70} (2004) 106008
  [arXiv:hep-th/0406208].

\bibitem{Forste:2005rs}
  S.~Forste, H.~P.~Nilles and A.~Wingerter,
  Phys.\ Rev.\ D {\bf 72} (2005) 026001
  [arXiv:hep-th/0504117].

\bibitem{Forste:2005gc}
  S.~Forste, H.~P.~Nilles and A.~Wingerter,
  arXiv:hep-th/0512270.


\bibitem{SO32:webseite}
H.~P.~Nilles, S.~Ramos-S\'anchez, P.~K.~S.~Vaudrevange and A.~Wingerter,
http://www.th.physik.uni-bonn.de/nilles/orbifolds

\bibitem{Georgi:1974my}
  H.~Georgi,
  AIP Conf.\ Proc.\  {\bf 23} (1975) 575.

\bibitem{Fritzsch:1974nn}
  H.~Fritzsch and P.~Minkowski,
  Annals Phys.\  {\bf 93} (1975) 193.

\bibitem{Nilles:2004ej}
  H.~P.~Nilles,
  arXiv:hep-th/0410160.

\bibitem{Angelantonj:1996uy}
  C.~Angelantonj, M.~Bianchi, G.~Pradisi, A.~Sagnotti and Y.~S.~Stanev,
  Phys.\ Lett.\ B {\bf 385} (1996) 96
  [arXiv:hep-th/9606169].

\bibitem{Kakushadze:1997wx}
  Z.~Kakushadze,
  Nucl.\ Phys.\ B {\bf 512} (1998) 221
  [arXiv:hep-th/9704059].

\bibitem{Lalak:1999bk}
  Z.~Lalak, S.~Lavignac and H.~P.~Nilles,
  Nucl.\ Phys.\ B {\bf 559} (1999) 48
  [arXiv:hep-th/9903160].

\bibitem{Blumenhagen:2005pm}
  R.~Blumenhagen, G.~Honecker and T.~Weigand,
  JHEP {\bf 0508} (2005) 009
  [arXiv:hep-th/0507041].


\bibitem{Gmeiner:2002es}
  F.~Gmeiner, S.~Groot Nibbelink, H.~P.~Nilles, M.~Olechowski and M.~G.~A.~Walter,
  Nucl.\ Phys.\ B {\bf 648} (2003) 35
  [arXiv:hep-th/0208146].


\bibitem{Ibanez:1987pj}
  L.~E.~Ibanez, J.~Mas, H.~P.~Nilles and F.~Quevedo,
  Nucl.\ Phys.\ B {\bf 301} (1988) 157.

\bibitem{Green:1987sp}
  M.~B.~Green, J.~H.~Schwarz and E.~Witten, {\em {Superstring Theory. Vol. 1: Introduction}}, App. 5.A.

\bibitem{Slansky:1981yr}
  R.~Slansky,
  Phys.\ Rept.\  {\bf 79}, 1 (1981).

\bibitem{Dixon:1986jc}
  L.~J.~Dixon, J.~A.~Harvey, C.~Vafa and E.~Witten,
  Nucl.\ Phys.\ B {\bf 274} (1986) 285.

\bibitem{Kac:1969xxx1}
V.~G. Kac, ``Automorphisms of finite order of semisimple lie algebras,'' {\em
  Func. Anal. Appl.} {\bf 3} (1969) 252.

\bibitem{Green:1984sg}
  M.~B.~Green and J.~H.~Schwarz,
  Phys.\ Lett.\ B {\bf 149} (1984) 117.

\bibitem{Witten:1984dg}
  E.~Witten,
  Phys.\ Lett.\ B {\bf 149} (1984) 351.

\bibitem{Dine:1987xk}
  M.~Dine, N.~Seiberg and E.~Witten,
  Nucl.\ Phys.\ B {\bf 289} (1987) 589.

\bibitem{Sagnotti:1992qw}
  A.~Sagnotti,
  Phys.\ Lett.\ B {\bf 294} (1992) 196
  [arXiv:hep-th/9210127].

\bibitem{Berkooz:1996iz}
  M.~Berkooz, R.~G.~Leigh, J.~Polchinski, J.~H.~Schwarz, N.~Seiberg and E.~Witten,
  Nucl.\ Phys.\ B {\bf 475} (1996) 115
  [arXiv:hep-th/9605184].

\bibitem{Kobayashi:1991rp}
  T.~Kobayashi and N.~Ohtsubo,
  Int.\ J.\ Mod.\ Phys.\ A {\bf 9} (1994) 87.

\bibitem{Casas:1991ac}
  J.~A.~Casas, F.~Gomez and C.~Munoz,
  Int.\ J.\ Mod.\ Phys.\ A {\bf 8} (1993) 455
  [arXiv:hep-th/9110060].

\bibitem{SVP}
The European String Vacuum Project website is located at\\
{\tt http://www.ippp.dur.ac.uk/$\sim$dgrell/svp/}.

\bibitem{Dienes:2006ut}
  K.~R.~Dienes,
  arXiv:hep-th/0602286.

\bibitem{Katsuki:1989bf}
  Y.~Katsuki, Y.~Kawamura, T.~Kobayashi, N.~Ohtsubo, Y.~Ono and K.~Tanioka,
  Nucl.\ Phys.\ B {\bf 341} (1990) 611.
 \; Y.~Katsuki, Y.~Kawamura, T.~Kobayashi, N.~Ohtsubo, Y.~Ono and K.~Tanioka,
DPKU-8904


\bibitem{Kobayashi:2004ud}
  T.~Kobayashi, S.~Raby and R.~J.~Zhang,
  Phys.\ Lett.\ B {\bf 593} (2004) 262
  [arXiv:hep-ph/0403065].

\bibitem{Kobayashi:2004ya}
  T.~Kobayashi, S.~Raby and R.~J.~Zhang,
  Nucl.\ Phys.\ B {\bf 704} (2005) 3
  [arXiv:hep-ph/0409098].

\bibitem{Buchmuller:2004hv}
  W.~Buchmuller, K.~Hamaguchi, O.~Lebedev and M.~Ratz,
  Nucl.\ Phys.\ B {\bf 712} (2005) 139
  [arXiv:hep-ph/0412318].

\bibitem{Buchmuller:2005jr}
  W.~Buchmuller, K.~Hamaguchi, O.~Lebedev and M.~Ratz,
  arXiv:hep-ph/0511035.


\bibitem{Ibanez:1986tp}
  L.~E.~Ibanez, H.~P.~Nilles and F.~Quevedo,
  Phys.\ Lett.\ B {\bf 187} (1987) 25.

\bibitem{Faraggi:1989ka}
  A.~E.~Faraggi, D.~V.~Nanopoulos and K.~j.~Yuan,
  Nucl.\ Phys.\ B {\bf 335} (1990) 347.

\bibitem{Braun:2005ux}
  V.~Braun, Y.~H.~He, B.~A.~Ovrut and T.~Pantev,
  Phys.\ Lett.\ B {\bf 618}, 252 (2005)
  [arXiv:hep-th/0501070].

\bibitem{Bouchard:2005ag}
  V.~Bouchard and R.~Donagi,
  Phys.\ Lett.\ B {\bf 633} (2006) 783
  [arXiv:hep-th/0512149].

\bibitem{Blumenhagen:2006ux}
  R.~Blumenhagen, S.~Moster and T.~Weigand,
  arXiv:hep-th/0603015.

\end{thebibliography}
\end{document}